\documentclass[12pt,epsfig]{article}
\usepackage{amssymb}
\usepackage{epsfig}
\newcommand{\beq}{\begin{equation}}
\newcommand{\eeq}{\end{equation}}
\newcommand{\bec}[1] {\begin{equation}\label{#1} }
\newcommand{\eec} {\end{equation} }
\begin{document}

\begin{flushright}
OSU-HEP-05-11\\
LPT-ORSAY-05-62\\
\end{flushright}
\begin{center}
{\Large {\bf Leptogenesis in Minimal Left-Right Symmetric Models} }\\
 \vspace*{1.5cm}
 {\large {\bf K.S.
 Babu\footnote{kaladi.babu@okstate.edu}  and A. Bachri}}
\footnote{abdelghafour.bachri@okstate.edu}\\
 \vspace*{0.5cm}
{\it Oklahoma Center for High Energy Physics}\\
{\it Department of Physics, Oklahoma State University\\
\vspace*{0.2cm}
Stillwater, OK~74078, USA}\\
\vspace*{0.7cm}
 {\large {\bf H. Aissaoui\footnote{aissaoui@th.u-psud.fr}}
}\\
 \vspace*{0.5cm}
{\it Laboratoire de Physique Th\'eorique, Universit\'e de Paris XI,
\\B\^atiment 210, 91405 Orsay Cedex,
France}\\
\vspace*{0.2cm}
{\it Laboratoire de Physique Math\'ematique et Physique Subatomique\\
Mentouri University, Constantine 25000 Algeria}\\
\end{center}
\begin{abstract}
We analyze lepton asymmetry induced in the decay of right--handed
neutrinos in a class of minimal left--right symmetric models. In
these models, which assume low energy supersymmetry, the Dirac
neutrino mass matrix is proportional to the charged lepton mass
matrix. As a result, lepton asymmetry is calculable in terms of 9
parameters, all measurable in low energy neutrino experiments. By
solving the Boltzmann equations numerically we show that adequate
baryon asymmetry is generated in these models. This however places
significant constraints on the light neutrino parameters.  We find
$\tan^2\theta_{12} \simeq m_1/m_2$ and $\theta_{13} = (0.01-0.07)$
for the neutrino oscillation angles, and
 $\beta \simeq \alpha + \pi/2$ for the Majorana phases.
\end{abstract}
\newpage
\section{Introduction}
The discovery of neutrino flavor oscillations in solar, atmospheric,
and reactor neutrino experiments \cite{neutrinoexperiments} may have
a profound impact on our understanding of the dynamics of the early
universe. This is because such oscillations are feasible only if the
neutrinos have small (sub--eV) masses, most naturally explained by
the seesaw mechanism \cite{seesaw}. This assumes the existence of
super-heavy right--handed neutrinos $N_i$ (one per lepton family)
with masses of order $(10^{8}-10^{14})~{\rm GeV}$. The light
neutrino masses are obtained from the matrix $ M_{\nu}\simeq
M_{D}M^{-1}_{R}{M_{D}}^{T}$ where $M_D$ and $M_R$ are respectively
the Dirac and the heavy Majorana right-handed neutrino (r.h.n) mass
matrices. The decay of the lightest right--handed neutrino $N_1$ can
generate naturally an excess of baryons over anti-baryons in the
universe \cite{Yanagida} consistent with cosmological observations.
The baryon asymmetry parameter is an important cosmological
observable constrained by Big Bang Nucleosynthesis and determined
recently with high precision by the WMAP experiment \cite{WMAP}:
\begin{equation}\label{wmap}
\eta_{B}\equiv \frac{n_B}{n_{\gamma}} = (6.5^{+0.4}_{-0.3} )\times
10^{-10}.
\end{equation}
The decay of $N_1$ can satisfy all three of the Sakharov conditions
\cite{zakharov} needed for successful generation of $\eta_{B}$ -- it
can occur out of thermal equilibrium, there is sufficient $C$ and
$CP$ violation, and there is also baryon number violation.  The last
condition is met by combining lepton number violation in the
Majorana masses of the right--handed neutrinos with $B+L$ violating
interactions of the Standard Model arising through the electroweak
sphaleron processes \cite{Kuzmin}.  A compelling picture emerges,
with the same mechanism explaining the small neutrino masses and the
observed baryon asymmetry of the universe. $\eta_{B}$ appears to be
intimately connected to  the observed neutrino masses and mixings.

A more careful examination of the seesaw structure would reveal
that, although there is an underlying connection, the light neutrino
mass and mixing parameters cannot determine the cosmological baryon
asymmetry, when the seesaw mechanism is implemented in the context
of the Standard Model (SM) gauge symmetry. It is easy to see this as
follows.  Without loss of generality one can work in a basis where
the charged lepton mass matrix and the heavy right--handed neutrino
Majorana mass matrix $M_R$ are diagonal with real eigenvalues.  The
Dirac neutrino mass matrix would then be an arbitrary complex $3
\times 3$ matrix with 18 parameters (9 magnitudes and 9 phases).
Three of the phase parameters can be removed by field redefinitions
of the left--handed lepton doublets and the right--handed charged
lepton singlets.  The neutrino sector will then have 18 ($=15+3$)
parameters.  9 combinations of these will determine the low energy
observables (3 masses, 3 mixing angles and 3 phases), while the
lepton asymmetry (and thus $\eta_{B}$) would depend on all 18
parameters, leaving it arbitrary.

In this paper we show that it is possible to quantitatively relate
$\eta_{B}$ to light neutrino mass and mixing parameters by
implementing the seesaw mechanism in the context of a class of
supersymmetric left--right models \cite{Mohapatra}.  We note that
unlike in the SM where the right--handed neutrinos appear as  rather
ad hoc additions, in the left--right symmetric models they are more
natural as gauge invariance requires their existence.  Supersymmetry
has the well--known merit of solving the gauge hierarchy problem.
With the assumption of a minimal Higgs sector, it turns out that
these models predict the relation for the Dirac neutrino mass matrix
\begin{eqnarray}
M_{D}=c
 \pmatrix{
 m_e &0 &0  \cr
  0 &m_{\mu}   &0 \cr
  0 &0 &m_{\tau}\cr},
\end{eqnarray}
where $c \simeq m_t/m_b$ is determined from the quark sector,
leaving only the Majorana mass matrix $M_R$ to be arbitrary.  3
phases in $M_R$ can be removed, leaving a total of 9 parameters
which determine both the low energy neutrino masses and mixings as
well as the baryon asymmetry.  It then becomes apparent that
$\eta_{B}$ is calculable in terms of the neutrino observables. There
have been other attempts in the literature to relate leptogenesis
with low energy observables \cite{LeptoConstraints, lepto2}. Such
attempts often make additional assumptions such as $M_D=M_{{\rm
up}}$ (which may not be fully realistic), or specific textures for
lepton mass matrices.

While a lot has been learned from experiments about the light
neutrino masses and mixings, a lot remains to be learned. Our
analysis shows that cosmology puts significant restrictions on the
light neutrino parameters.  Successful baryogenesis requires within
our model that three conditions be satisfied: $\tan^2\theta_{12}
\simeq m_1/m_2$, $\beta \simeq \alpha + \pi/2$ and $\theta_{13} =
(0.01-0.07)$.  Here $\theta_{12}$ and $\theta_{13}$ are elements of
the neutrino mixing matrix, $m_i$ are the light neutrino mass
eigenvalues and $\alpha, \beta$ are the Majorana phases entering in
the amplitude for neutrinoless double beta decay.  Future neutrino
experiments will be able to either confirm or refute these
predictions.

The rest of the paper is organized as follows.  In Sec. 2 we review
briefly the minimal left--right symmetric model.  In Sec. 3 we
analyze leptogenesis in this model.  Here we derive constraints
imposed on the model from the requirement of successful
leptogenesis.  In Sec. 4 we calculate the lepton asymmetry parameter
$\varepsilon_1$ generated in the model in $N_1$ decay.  Sec. 5
summarizes the relevant Boltzmann equations needed for computing the
baryon asymmetry parameter.  Sec. 6 provides our numerical results
for $\eta_B$.  Finally, in Sec. 7 we conclude.

\section{Brief review of the minimal left-right symmetric model}

Let us briefly review the basic structure of the minimal SUSY
left--right symmetric model developed in Ref. \cite{Mohapatra}.  The
gauge group of the model is $SU(3)_C \times SU(2)_L\times
SU(2)_R\times U(1)_{B-L}$. The quarks and leptons are assigned to
the gauge group as follows. Left--handed quarks and leptons ($Q,L$)
transform as doublets of $SU(2)_L$ [$Q(3,2,1,1/3)$  and
$L(1,2,1,-1)$], while the right--handed ones ($Q^c,L^c$) are
doublets of $SU(2)_R$ [$Q^c(3^*,1,2,-1/3)$ and $L^c(1,1,2,1)$]. The
Dirac masses of fermions arise through their Yukawa couplings to a
Higgs bidoublet $\Phi(1,2,2,0)$. The $SU(2)_R\times U(1)_{B-L}$
symmetry is broken to $U(1)_Y$ by the VEV ($v_R$) of a $B-L=-2$
triplet scalar field $\Delta^c(1,1,3,-2)$. This triplet is
accompanied by a left--handed triplet $\Delta(1,3,1,2)$ (along with
$\bar{\Delta}$ and $\bar{\Delta^c}$ fields, their conjugates to
cancel  anomalies). These fields also couple to the leptons and are
responsible for inducing large Majorana masses for the $\nu_R$. An
alternative to these triplet Higgs fields is to use $B-L=\pm1$
doublets $\chi(1,2,1,-1) $  and $\chi^c(1,1,2,1)$,  along with their
conjugates $\bar{\chi}$ and $\bar{\chi^c}$.  In this case
non--renormalizable operators will have to be invoked to generate
large $\nu_R$ Majorana masses. For definiteness we shall adopt the
triplet option, although our formalism allows for the addition of
any number of doublet Higgs fields as well. The superpotential
invariant under the gauge symmetry involving the quark and lepton
fields is
\begin{eqnarray}
W  =  {\bf Y}_q Q^T \tau_2 \Phi \tau_2 Q^c + {\bf Y}_l L^T \tau_2
\Phi \tau_2 L^c
   +   ( {\bf f} L^T i\tau_2 \Delta L + {\bf f}_c
{L^c}^T i\tau_2 \Delta^c L^c)~. \label{sup}
\end{eqnarray}

Under left--right parity symmetry, $Q \leftrightarrow Q^{c*}, L
\leftrightarrow L^{c*}, \Phi \leftrightarrow \Phi^\dagger$,
$\Delta\leftrightarrow \Delta^{c*}$, along with $W_{SU(2)_L}
\leftrightarrow W^*_{SU(2)_R}$, $W_{B-L} \leftrightarrow W^*_{B-L}$
and $\theta \leftrightarrow \bar{\theta}$. As a consequence, ${\bf
Y}_q = {\bf Y}_q^\dagger$, ${\bf Y}_l = {\bf Y}_l^\dagger$, and
${\bf f} = {\bf f}^*_c$ in Eq. (\ref{sup}).\footnote{We do not
explicitly use these relations.} It has been shown in Ref.
\cite{Mohapatra} that the hermiticity of the Yukawa matrices (along
with the parity constraints on the soft SUSY breaking parameters)
helps to solve the supersymmetric CP problem that haunts the MSSM.

Below $v_R$, the effective theory is the MSSM with its $H_u$ and
$H_d$ Higgs multiplets.\footnote{The right-handed gauge bosons have
masses of order $v_R\sim 10^{14}~{\rm GeV}$ and thus play no
significant role in cosmology at $T\sim M_1 \ll v_R$.} These are
contained in the bidoublet $\Phi$ of the SUSY left-right model, but
in general they can also reside partially in other multiplets having
identical quantum numbers under the MSSM symmetry (such as the
$\chi, ~\overline{\chi}$ doublet Higgs fields alluded to earlier).
Allowing for such a possibility, the superpotential of Eq.
(\ref{sup}) leads to the relations for the MSSM Yukawa coupling
matrices
\begin{eqnarray}\label{gammarelation}
{\bf Y}_u~=\gamma {\bf Y}_d, ~~~~~~~~~~~~~~~{\bf
Y}_{\ell}~=\gamma{\bf Y}_{\nu^D}\label{yuk}~.
\end{eqnarray}
These relations have been called up--down unification
\cite{Mohapatra}. Here, the first relation of Eq.
(\ref{gammarelation}) implies $\frac{m_t}{m_b}\simeq \gamma
\tan\beta \equiv c$ where $\gamma$ is a parameter characterizing how
much of $H_u$ and $H_d$ of MSSM are in the bidoublet $\Phi$. The
case of $H_{u,d}$ entirely in $\Phi$ will correspond to $\gamma = 1$
and $\tan\beta = m_t/m_b$. At first sight the first of the relations
in Eq. (\ref{yuk}) might appear phenomenologically disastrous since
it leads to vanishing quark mixings and unacceptable quark mass
ratios. It was shown in the first paper of Ref. \cite{Mohapatra}
that including the one--loop diagrams involving the gluino and the
chargino and allowing for a flavor structure for the soft SUSY
breaking $A$ terms, there exists a large range of parameters (though
not the entire range possible in the usual MSSM) where correct quark
mixings as well as masses can be obtained consistent with flavor
changing constraints.

It is the second of Eq. (\ref{yuk}) that concerns us here. This
relation would lead to $M_D = c M_l$, with $c \simeq m_t/m_b$. One
can thus go to a basis where the charged lepton and the Dirac
neutrino mass matrices are simultaneously diagonal.  The heavy
Majorana mass matrix $M_R = {\bf f} v_R$ will then be a generic
complex symmetric matrix.  After removing three phases in $M_R$ by
field redefinitions, we are left with 9 parameters (6 magnitudes and
3 phases) which determine the light neutrino spectrum as well as the
heavy neutrino spectrum. This in turns fixes the lepton asymmetry.
The consequences of such a constrained system for leptogenesis will
be analyzed in the next section.

In principle the $\Delta (1,3,1,+2)$ Higgs field can also acquire a
small VEV $\lesssim {\cal O}({\rm eV})$ \cite{type2seesaw}. In this
case the seesaw formula would be modified, as will the calculation
of the lepton asymmetry \cite{type2seesaw}. We will assume such type
II seesaw contributions proportional to $\left
\langle\Delta\right\rangle$ are zero in our analysis. This is
consistent with the models of Ref. \cite{Mohapatra}. Leptogenesis in
the context of more general left-right symmetric models has been
analyzed in Ref. \cite{LRpapers}.
\section{Leptogenesis in left-right symmetric framework}
The $SU(2)_{R}\times U(1)_{B-L}$ symmetry is broken down to $U(1)_Y$
by the VEV $\left\langle\Delta^c\right\rangle=v_R\sim 10^{14}~{\rm
GeV}$. At least some of the right-handed neutrinos have masses below
$v_R$. We thus focus on the neutrino Yukawa coupling in the context
of MSSM. The $SU(2)_{L}\times U(1)_Y$ invariant Yukawa interactions
are contained in the MSSM superpotential
\begin{eqnarray}
W = l H_d {\bf Y}_{\ell} ~e^c + l H_u {\bf Y}_{\nu^D} ~\nu^c +
\frac{1}{2}\underbrace{ \nu^{cT}CM_R \nu^c},
\end{eqnarray}
where $l$ stands for the left-handed lepton doublet, and
$(e^c,\nu^c)$ denote the conjugates of the right-handed charged
lepton and the right--handed neutrino fields respectively. $H_u$,
$H_d$ are the MSSM Higgs fields with VEVs $v_u$, $v_d$. $M_l={\bf
Y}_{\ell}~v_d$, $M_D={\bf Y}_{\nu^D}~ v_u$ and $M_R$ are
respectively the charged lepton, the Dirac neutrino, and the
Majorana r.h.n mass matrices. Then one can generate light neutrino
masses by the seesaw mechanism \cite{seesaw}
\begin{eqnarray}\label{seasaw}
M_{\nu}=-M_{D}M^{-1}_{R}{M_{D}}^{T}.
\end{eqnarray}
There is mixing among generations in both $M_R$ and $M_D$, the light
neutrino mixing angles will depend on both of these mixings. Within
the SM or MSSM where $M_D$ is an arbitrary matrix, the structure of
the right-handed neutrino mass matrix can not be fully determined
even if the light matrix $M_{\nu}$ were to be completely known from
experiments. As noted in Sec. 2, in the minimal version of the
left-right symmetric model one has
\begin{eqnarray}\label{LeftRightEq}
M_{D}=c M_{l}=c~diag(m_{e},m_{\mu},m_{\tau})
\end{eqnarray}
where $c\simeq\frac{m_t}{m_b}$. Here we have already gone to a basis
where the charged lepton mass matrix is diagonalized. In the three
family scenario, the relations between the flavor eigenstates
$(\nu_e,\nu_{\mu},\nu_{\tau})$ and the mass eigenstates
$(\nu_1,\nu_2,\nu_3)$ can be expressed in terms of observables as
\begin{equation}\label{lightneutrinomatrix}
M_{\nu} = U^{\ast} M_{\nu}^{diag} U^{\dagger},
\end{equation}
where $M_{\nu}^{diag}\equiv diag(m_1,m_2,m_3)$, with $m_i$ being the
light neutrinos masses and $U$ being the $3\times 3$ mixing matrix
which we write as $U = U_{PMNS}.P$. We parameterize $U_{PMNS}$
\cite{Maki} as
\begin{eqnarray}
U_{PMNS} & = &
 \pmatrix{
 U_{e1}&U_{e2} &U_{e3} \cr
  U_{\mu 1}&U_{\mu 2}&U_{\mu 3} \cr
 U_{\tau 1} &U_{\tau 2} &U_{\tau 3} \cr}\nonumber\\
 & = &\pmatrix{
 c_{12}c_{13}&s_{12}c_{13} &s_{13}e^{-\imath\delta} \cr
  -s_{12}c_{23}-c_{12}s_{13}s_{23}e^{\imath\delta}&c_{12}c_{23}-s_{12}s_{13}s_{23}
  e^{\imath\delta} &c_{13}s_{23} \cr
s_{12}s_{23}-c_{12}s_{13}c_{23}e^{\imath\delta}
&-c_{12}s_{23}-s_{12}s_{13}c_{23}e^{\imath\delta}&c_{13}c_{23} \cr}
\end{eqnarray}
where $c_{ij}\equiv \cos\theta_{ij}$, $s_{ij}\equiv \sin\theta_{ij}$
and $\delta$ is the Dirac CP violating phase which appears in
neutrino oscillations. The matrix $P$ contains two Majorana phases
unobservable in neutrino oscillation, but relevant to neutrinoless
double beta decay \cite{valle}:
\begin{eqnarray}
 P =
 \pmatrix{
 e^{\imath \alpha}&0 &0 \cr
  0&e^{\imath \beta} &0 \cr
 0 &0 &1 \cr}.
\end{eqnarray}

Combining Eq. (\ref{lightneutrinomatrix}) with the seesaw formula of
Eq. (\ref{seasaw}) and solving for the right-handed neutrino mass
matrix we find
\begin{eqnarray}\label{MRMNS}
M_R & = & c^2 M_l \,M^{-1}_{\nu}M_l\nonumber\\
    & = &\frac{c^2m^2_\tau}{m_1}\pmatrix{
 \frac{m_e}{m_\tau}&0 &0 \cr
  0&\frac{m_\mu}{m_\tau}&0 \cr
 0 &0 &1 \cr}U_{PMNS}P^2\pmatrix{
 1&0 &0 \cr
  0&\frac{m_1}{m_2}&0 \cr
 0 &0 &\frac{m_1}{m_3} \cr}U^T_{PMNS}\pmatrix{
 \frac{m_e}{m_\tau}&0 &0 \cr
  0&\frac{m_\mu}{m_\tau}&0 \cr
 0 &0 &1 \cr}.
\end{eqnarray}
This enables us to establish a link between high scale parameters
and low scale observables.

We define a small expansion parameter $\epsilon$ as
$$\epsilon=\frac{m_\mu}{m_\tau}\simeq0.059,$$
in terms of which we have
\begin{eqnarray}\label{parameters1}
m_e = a_e \epsilon^3 m_\tau, \,\,\frac{m_1}{m_3}=a_{13}\epsilon,
\,\,\theta_{13}=t_{13}\epsilon, \,\,
\theta_{23}=\frac{\pi}{4}+t_{23}\epsilon.
\end{eqnarray}
Here $a_e$, $a_{13}$, $t_{13}$ and $t_{23}$ are $\lesssim {\cal
O}(1)$ parameters with $a_e=1.400$. These expansions follow from low
energy data assuming the picture of hierarchical neutrino masses.

We find that the requirement of generating adequate baryon asymmetry
places significant constraints on the neutrino mixing parameters.
Specifically, the following expansions
\begin{eqnarray}\label{parameters2}
\frac{m_1}{m_2}=\tan^2\theta_{12}+a_{12}\epsilon \,\,\mbox{and} \,\,
\beta=\alpha+\frac{\pi}{2}+b\epsilon,
\end{eqnarray}
where $a_{12}$ and $b$ are $\lesssim {\cal O}(1)$ parameters are
required. To see this, we note that the CP asymmetry parameter
$\epsilon_1$ generated in the decay of $N_1$ is too small, of order
$\varepsilon_1\sim\frac{\epsilon^6}{8\pi}\sim 2\times 10^{-9}$ if
$a_{12}$ or $b$ are much greater than 1. This is because the heavy
neutrino masses would be strongly hierarchical in this case, $M_1:
M_2: M_3\sim \epsilon^6: \epsilon^2: 1$. This can be altered to a
weak hierarchy $M_1: M_2: M_3\sim \epsilon^4: \epsilon^2: 1$ by
observing that the elements of the 2-3 block of $M_R$ of Eq.
(\ref{MRMNS}) are all proportional to
$\{\frac{m_1}{m_2}e^{2i\beta}\cos^2\theta_{12}+e^{2i\alpha}\sin^2\theta_{12}\}$
and by demanding this quantity to be of order $\epsilon$. Eq.
(\ref{parameters2}) is just this condition. $\varepsilon_1\sim
\frac{\epsilon^4}{8\pi}\sim 10^{-6}$ in this case, which can lead to
acceptable baryon asymmetry, as we show.

An immediate consequence of Eq. (\ref{parameters2}) is that
neutrinoless double beta decay is suppressed in the model. The
effective mass relevant for this decay is found to be
\begin{equation}
m_{\beta\beta}=|\sum_i U^2_{ei}m_i|\simeq |m_2e^{2i\alpha}\epsilon
(a_{12}c^2_{12}-2ib s^2_{12})+m_3 s^2_{13}e^{-2i\delta}|.
\end{equation}
This is of the order $m_3\epsilon^2\sim 10^{-4}$ eV, which would be
difficult to measure. This amplitude is small because of a
cancelation between the leading contributions proportional to $m_1$
and $m_2$ (see Eq. (\ref{parameters2})).

In terms of these expansions, the r.h.n mass matrix becomes
\begin{eqnarray}\label{MR}
M_R = \pmatrix{ A_{11}\epsilon^5&A_{12}\epsilon^3 &A_{13}\epsilon^2
\cr
  A_{12}\epsilon^3&A_{22}\epsilon^2&A_{23}\epsilon \cr
 A_{13}\epsilon^2 &A_{23}\epsilon &A_{33} \cr},
\end{eqnarray}
where
\begin{eqnarray}
A_{11}&= &\frac{M_{\circ}\epsilon\, a^2_e e^{2\imath\alpha}\cos2\theta_{12}}{\cos^2\theta_{12}}\nonumber\\
A_{12}&= &-\frac{M_{\circ}\epsilon\, a_e e^{2\imath\alpha}\tan\theta_{12}}{\sqrt{2}}\nonumber \\
A_{13}&= &\frac{M_{\circ}\epsilon \,a_e e^{2\imath\alpha}\tan\theta_{12}}{\sqrt{2}} \nonumber\\
A_{22}&= &
           \frac{M_{\circ}\epsilon}{2}\left\{a_{13}- a_{12}e^{2\imath\alpha}\cos^2\theta_{12}
           - 2\imath b e^{2\imath\alpha}\sin^2\theta_{12}+
           2e^{\imath(2\alpha+\delta)}t_{13}\tan\theta_{12}\right\}\nonumber\\
A_{23}& =&
           \frac{M_{\circ}\epsilon}{2}\left\{a_{13}+ a_{12}e^{2\imath\alpha}\cos^2\theta_{12}
           + 2\imath b e^{2\imath\alpha}\sin^2\theta_{12}\right\}\nonumber\\
A_{33}&= &-M_{\circ}\epsilon\,
         e^{2\imath\alpha}\left\{t_{13}e^{\imath\delta}\tan\theta_{12}+\imath b\sin^2\theta_{12}
        -\frac{a_{13}e^{-2\imath\alpha}}{2}+\frac{a_{12}\cos^2\theta_{12}}{2}\right\}.
\end{eqnarray}
Here we defined $M_{\circ} = \frac{c^2 m^2_{\tau}}{m_1}$. This
hierarchical mass matrix is diagonalized by a series of rotations
$U_1$, $U_2$ and $U_3$ such that;
\begin{eqnarray}
(KU_3U_2U_1)M_R(KU_3U_2U_1)^T =\pmatrix{ |M_1|&0 &0 \cr
  0&|M_2|&0 \cr
 0 &0 &|M_3| \cr}
\end{eqnarray}
where $K=diag(k_1,k_2,k_3)$ with $k_i=e^{-\imath \phi_i /2}$ being
phase factors which make each r.h.n masses $M_i$ real,
$M_i=|M_i|e^{\phi_i}$. $V=(KU_3U_2U_1)^T$ is the matrix that
diagonalizes $M_R$. The unitary matrix $U_1$ is given by
\begin{eqnarray}
U_1 = \pmatrix{ 1&0&-\frac{A_{13}}{A_{33}}\epsilon^2 \cr
  0&1&0 \cr
 \frac{A^{\star}_{13}}{A^{\star}_{33}}\epsilon^2 &0 &1 \cr}.
\end{eqnarray}
Similarly, $U_2$ and $U_3$ are unitary matrices with off-diagonal
entries given by
\begin{equation}
(U_2)_{23}=-\frac{A_{23}}{A_{33}}\epsilon~~,~~~~(U_3)_{12}=-\frac{\left(A_{12}-\frac{A_{13}A_{23}}{A_{33}}\right)\epsilon}
{A_{22}-\frac{A^2_{23}}{A_{33}}} .
\end{equation}
The mass eigenvalues are found to be
\begin{eqnarray}
M_1&=&M_{\circ}k_1^2\epsilon^5\left(2a_{13}a_e^2 e^{2\imath
\alpha}\sin^2\theta_{12}\right) \nonumber\\
&\times& \left(2t^2_{13}
        e^{2\imath(\alpha+\delta)}\sin^2\theta_{12}\right.
        +\left.(a_{12}+2\imath b+(a_{12}-2\imath
        b)\cos2\theta_{12})a_{13}\cos^2\theta_{12}\right)^{-1}\nonumber\\
\nonumber\\
M_2&=&M_{\circ}k_2^2\epsilon^3 e^{2\imath \alpha}
\left(a_{13}(a_{12}+2\imath b+ (a_{12}-2\imath
b)\cos2\theta_{12})+2t^2_{13}e^{\imath(\delta+\alpha)}\tan^2\theta_{12}\right)\nonumber\\
&\times& \left(-a_{13}+\imath b
e^{2\imath\alpha}+e^{2\imath\alpha}(a_{12}\cos^2\theta_{12}-\imath b
\cos2\theta_{12})+2e^{\imath\delta}t_{13}\tan\theta_{12}
\right)^{-1}\nonumber\\
\nonumber\\
M_3&=&\frac{M_{\circ}k^2_3\epsilon}{2} \,\left(a_{13}-\imath b
e^{2\imath
            \alpha}-e^{2\imath\alpha}(a_{12}\cos^2\theta_{12}-\imath b
               \cos2\theta_{12}+ 2t_{13}e^{\imath
               \delta}\tan\theta_{12}) \right).
\end{eqnarray}
We use these results in the next section to determine
$\varepsilon_1$.
\section{CP violation and lepton asymmetry} Now that we
have developed our framework, we can turn attention to the
evaluation of the CP asymmetry $\varepsilon_{1}$ generated in the
decay of the lightest r.h.n $N_1$. This arises from the interference
between the tree-level and one-loop level decay
amplitudes.\footnote{We will assume $M_1\ll M_2< M_3$. In this case,
even if the heavier right-handed neutrinos $N_2$ and $N_3$ produce
lepton asymmetry, it is usually erased before the decay of $N_1$.}
In a basis where the r.h.n mass matrix is diagonal and real, the
asymmetry in the decay  of $N_i$ is given by \cite{RouletVissani}
\begin{eqnarray}\label{CP}
    \varepsilon_i=-\frac {1}{8\pi\upsilon^2\,(M^{\dagger}_D M_D)_{ii}}  \,
    \sum_{j=2,3} \mbox{Im} [(M^{\dagger}_D M_D)_{ij}]^2\,
    \left[f\left(\frac{M^2_j}{M^2_i}\right)+g\left(\frac{M^2_j}{M^2_i}\right)\right]
\end{eqnarray}
where $f(x)$ and $g(x)$ represent the contributions from vertex and
self energy corrections respectively. For the case of the
non-supersymmetric standard model with right-handed neutrinos, these
functions are given by~\cite{RouletVissani}
\begin{eqnarray}
 \label{EQ-ep1-function-nonSUSY}
  f_{\rm non-SUSY}(x)
  =
  \sqrt{x}
  \left[
   -1
   +
   (x+1)\ln\left(1+\frac{1}{x}\right)
   \right]
   \,,
   \qquad
   g_{\rm non-SUSY}(x)
   =
   \frac{\sqrt{x}}{x-1}
   \,,
\end{eqnarray}
while for the case of MSSM plus right-handed neutrinos, they are
given by
\begin{eqnarray}
 \label{EQ-ep1-function}
  f_{\rm SUSY}(x)
  =
  \sqrt{x}\,
  \ln\left(1+\frac{1}{x}\right)
  \,,
  \qquad
  g_{\rm SUSY}(x)
  =
  \frac{2 \sqrt{x}}{x-1}
  \,.
\end{eqnarray}
Here $\upsilon$ is the SM Higgs doublet VEV, $\upsilon\simeq 174$
GeV. For the case of MSSM, $\upsilon$ in Eq. (\ref{CP}) is replaced
by $\upsilon \sin\beta$. Hereafter, for definiteness in the
numerical evaluation of the Boltzmann equations, we assume the SM
scenario. However, our result should be approximately valid for the
MSSM case as well.\footnote{The function $f+g$ in MSSM is twice as
big compared to the SM. However this is compensated by the factor
$\frac{1}{g_{\ast}}$ that appears in $\eta_{B}$ which in MSSM is
half of the SM value.} Assuming a mass hierarchy $M_1\ll M_2< M_3$
in the right-handed neutrino sector i.e., ($x\gg 1$), which is
realized in our model, see Eq. (\ref{MR}), the above formula is
simplified to the following one:
\begin{eqnarray}\label{CPTOTAL}
\varepsilon_1=-\frac{3}{16\pi\upsilon^2(M_D^{\dag} M_D)_{11}}
\sum_{k=2,3}\,\mbox{Im}[\,(M_D^{\dag} M_D)_
{1k}^2\,]\,\frac{M_1}{M_k}.
\end{eqnarray}
$\varepsilon_1$ depends on the (1,1), (1,2) and (1,3) entries of
$M^{\dagger}_D M_D$. These quantities can be related to the light
neutrino mass and mixing parameters measurable in low energy
experiments. In the basis where $M_R$ is diagonal, these elements
are
\begin{eqnarray}
(M^{\dagger}_D M_D)_{11}& = &
(c\,m_{\tau})^2\left(V_{31}V_{31}^{\ast}+
V_{21}V_{21}^{\ast}\epsilon^2 + a_e^2 V_{11}V_{11}^{\ast}\epsilon^6\right)\nonumber\\
 (M^{\dagger}_D M_D)_{12}& = &
(c\,m_{\tau})^2\left(V_{31}V_{32}^{\ast}+
V_{21}V_{22}^{\ast}\epsilon^2 + a_e^2 V_{11}V_{12}^{\ast}\epsilon^6\right)\nonumber\\
(M^{\dagger}_D M_D)_{13}& = &
(c\,m_{\tau})^2\left(V_{31}V_{33}^{\ast}+
V_{21}V_{23}^{\ast}\epsilon^2 + a_e^2
V_{11}V_{13}^{\ast}\epsilon^6\right),
\end{eqnarray}
where $V=K U_3 U_2 U_1$ is the unitary matrix diagonalizing $M_R$.
Straightforward calculations give, to leading order in $\epsilon$,
\begin{eqnarray}
(M^{\dagger}_D M_D)_{11}& = &
8a^2_ec^2m^2_{\tau}\epsilon^4\,\cos^2\theta_{12}\sin^2\theta_{12}
(a^2_{13}+t^2_{13}\tan^2\theta_{12})\nonumber\\
& \times & 1/
\left\{8t^4_{13}\sin^4\theta_{12}+32a_{13}t^2_{13}b\cos^2\theta_{12}\sin^4\theta_{12}\sin
2(\alpha+\delta)\right.\nonumber\\
&+&a_{13}\cos^4\theta_{12}[4a_{13}(a^2_{12}-b^2)\cos
2\theta_{12}+a_{13}(a^2_{12}+4b^2)(3+\cos
4\theta_{12})\nonumber\\
&+& \left.16a_{12}t^2_{13}\sin^2\theta_{12}\cos 2(\alpha+\delta) ] \right\}\\
\nonumber\\
(M^{\dagger}_D M_D)_{12}^2 & = &
2a^2_ec^4m^4_{\tau}\epsilon^6\tan^2\theta_{12} e^{-\imath
(\phi_1-\phi_2)}e^{-2\imath(2\alpha+\delta)}\left\{4(a^2_{13}-t^2_{13})\cos
2\theta_{12}-2t_{13}\sin
2\theta_{12}\right.\nonumber\\
&&\left.(2a_{13}e^{\imath(2\alpha+\delta)}-(a_{12}+2\imath
b)e^{-\imath\delta})+4(a^2_{13}+t^2_{13})+t_{13}\sin 4\theta_{12}e^{-\imath\delta}(a_{12}-2\imath b)\right\}^2\nonumber\\
& \times & 1/\left\{[\imath b
e^{\imath\delta}-a_{13}e^{-\imath(2\alpha+\delta)}+a_{12}e^{-\imath\delta}\cos^2\theta_{12}
-\imath b e^{-\imath\delta}\cos
2\theta_{12}+2t_{13}\tan\theta_{12}]^2\right.\nonumber\\
& \times &\left. [3a_{12}a_{13}-2\imath
a_{13}b+4t^2_{13}e^{-2\imath(\alpha+\delta)}+4\cos
2\theta_{12}(a_{12}a_{13}-t^2_{13}e^{-2\imath(\alpha+\delta)})\,\,+\right.\nonumber\\
& &\left.a_{13}(a_{12}+2\imath b)\cos
4\theta_{12}]^2\right\}\\
\nonumber\\
(M^{\dagger}_D M_D)_{13}^2 &=&2a^2_e c^4
m^4_{\tau}\epsilon^4\sin^2\theta_{12}e^{-\imath
(\phi_1-\phi_3)}(a_{13}\cos\theta_{12}+e^{-\imath
(2\alpha+\delta)}t_{13}\sin\theta_{12})^2\nonumber\\
&\times &1/ \left\{a_{13}\cos^2\theta_{12}(a_{12}-2\imath b
+(a_{12}+2\imath b)\cos
2\theta_{12})+2t^2_{13}\sin^2\theta_{12}e^{-2\imath
(\alpha+\delta)}\right\}^2
\end{eqnarray}
These analytical expressions have been checked numerically. In
Figure (1) we have plotted $|\varepsilon_1|$ as function of
$\theta_{13}$ for fixed values of other observables. The solid line
in Fig (1) which corresponds to the exact numerical evaluation
agrees very well with the dashed line corresponding to the
analytical expressions.

From Figure (1), it is apparent that $\theta_{13}$ is constrained in
the model from cosmology. If $\varepsilon_1 < 1.3\times 10^{-7}$,
the induced baryon asymmetry would be too small to explain
observations. As can be seen from Figure (1), $\theta_{13}$ should
lie in the range $0.01-0.07$ for an acceptable value of
$\varepsilon_1$. This result does not change very much with
variations in the other input parameters.
\begin{figure}
\begin{center}
\includegraphics*[bb= 76 2 505 254,width=6in]{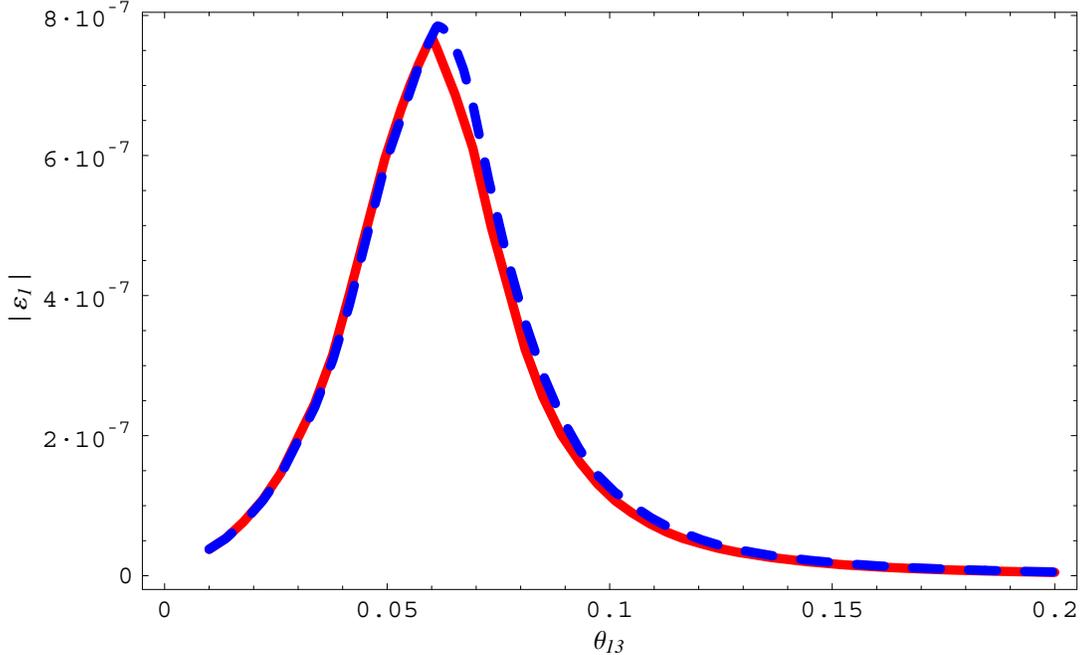}
\caption{Plots for
 CP asymmetry parameter $\varepsilon_1$ using analytical (dotted) and numerical (solid) results as
a function of the neutrino oscillation angle $\theta_{13}$. The
input parameters used are $a_{12}=1$, $b=1$, $\Delta m^2_{\odot}=
2.5 \times 10^{-5}\,{\rm eV}^2$, $\Delta m^2_{a}= 5.54\times 10^{-3}
\,{\rm eV}^2$
 and $\{\delta, \alpha\}=\{\pi/4, \pi/4\}$. Our model requires
$|\varepsilon_1|\gtrsim1.3\times 10^{-7}$ to successfully generate
an adequate number for the BA. This criterium happens to be
satisfied only in the region for which $0.01\lesssim
\theta_{13}\lesssim 0.07$, this interval is not too sensitive to
variations in the input parameters.}
\end{center}
\end{figure}
\vspace*{-0.25in}

Electroweak sphaleron processes \cite{Kuzmin} will convert the
induced lepton asymmetry to baryon asymmetry. The ratio of baryon
asymmetry to entropy $Y_B$ is related to the lepton asymmetry
through the relation \cite{YBL}:
\begin{eqnarray}\label{conversion}
    Y_B =C\,Y_{B-L}= \frac{C}{C-1}Y_L
\end{eqnarray}
where $C=\frac{8N_f+4N_{\varphi}}{22N_f+13N_{\varphi}}$, $N_f=3$ and
$N_{\varphi}=1, 2$ in the case of the SM and MSSM respectively. In
either case $C\sim\frac{1}{3}$. In Eq. (\ref{conversion}),
$Y_B=\frac{n_B}{s}$ with $s=7.04~n_{\gamma}$.

There has been considerable interest in obtaining approximate
analytical expression for baryon asymmetry \cite{efficiency,
recentefficiency}. In order to estimate this, the dilution factor,
often referred to as the efficiency factor $\kappa$ that takes into
account the washout processes (inverse decays and lepton number
violating scattering) has to be known. As an example, $\kappa=(2\pm
1)\times 10^{-2}\left( \frac{0.01 \mbox{ eV}}{\widetilde{m}_1}\right
)^{1.1\pm 0.1}$ has been suggested in Ref. \cite{efficiency} from
which $ \eta_{B}\simeq 0.96\times 10^{-2}\varepsilon_{N_1}\kappa$
has been calculated. In our work we solve the coupled Boltzmann
equations numerically to estimate the baryon asymmetry without
referring to the efficiency factor.
\section{Boltzmann equations}
In this section we set up the Boltzmann equations for computing the
baryon asymmetry $\eta_B$ generated through the out of equilibrium
decay of $N_1$. In our model the right-handed neutrino masses are
not independent of the CP asymmetry parameter $\varepsilon_1$.
Therefore a self consistent analysis within the model is required.

In the early universe, at temperature of order $N_1$ mass, the main
thermal processes which enter in the production of the lepton
asymmetry are the
decay of the lightest r.h. neutrino,\footnote{%
In our analysis we stick to the case where the asymmetry is due only
to the decay of the lightest r.h. neutrino $N_{1}$.} its inverse
decay, and the lepton number violation scattering, $\Delta L=1$
Higgs exchange plus $\Delta L=2$ r.h.n exchange \cite{Luty}. The
production of the lepton asymmetry via the decay of the r.h.n is an
out-of-equilibrium process which is most efficiently treated by
means of the Boltzmann equations (BE).

The first BE which describes the evolution of the abundance of the
r.h. neutrino and which corresponds to the source of the asymmetry
is given by\footnote{In this section we follow the notation of the
first paper of Ref. \cite{LeptoConstraints} to which we refer the
reader for further details.}
\begin{equation}
{\frac{dY_{N_{1}}}{dz}}=-{\frac{z}{H s(z)}}\left({\frac{Y_{N_{1}}}{{Y_{N}^{eq}}%
}}-1\right)\left(\gamma _{_{D_{1}}}+\gamma _{_{S_{1}}}\right),
\label{be1}
\end{equation}
where $z=\frac{M_{1}}{T}$. Here $s(z)$ is the entropy density and
$\gamma
_{_{D_{1}}},\gamma _{_{S_{1}}}$ are the interaction rates for the decay and $%
\Delta L=1$ scattering contributions, respectively.

The second BE relevant to the lepton asymmetry is given by
\begin{equation}
{\frac{dY_{B-L}}{dz}}=-{\frac{z}{s(z)H(M_{1})}}\left[\varepsilon
_{1}\gamma
_{_{D_{1}}}\left({\frac{Y_{N_{1}}}{Y_{N}^{eq}}}-1\right)+\gamma _{_{W}}{\frac{Y_{B-L}}{%
Y_{L}^{eq}}}\right],  \label{be2}
\end{equation}
where $\varepsilon _{1}$ is the CP violation parameter given by Eq. (\ref{CP}%
) and $\gamma _{_{W}}$ is the washout factor which is responsible
for damping of the produced asymmetry, see Eq. (\ref{gammaW}) below.
In Eqs. (\ref{be1}) and (\ref{be2}), $Y_{i}^{eq}$ is the equilibrium
number density of a particle species $i$, which has a mass $m_{i}$,
given by
\begin{equation}
Y_{i}^{eq}(z)={\frac{45}{4\pi ^{4}}}{\frac{g_{i}}{g_{\ast }}}\left( {\frac{%
m_{i}}{M_{1}}}\right) ^{2}z^{2}K_{2}\left(
{\frac{m_{i}z}{M_{1}}}\right) \ ,
\end{equation}
where $g_{i}$ is the particle internal degree of freedom ($g_{_{N_{i}}}=2$, $%
g_{_{\ell }}=4$). At temperatures far above the electroweak scale
one has $g_*\simeq 106.75$ in the standard model, and $g_*\simeq
 228.75$ in MSSM. $H$, the Hubble parameter evaluated at $z=1$, and $s(z)$, the entropy
density, are given by
\begin{equation}
\label{Hs} H = \sqrt{\frac{4 \pi^3 g_{\ast}}{45}}\,
\frac{M_1^2}{M_P}\quad,\quad s(z) =\frac{2 \pi^2 g_{\ast}}{45}
\frac{M_1^3}{z^3}\ ,
\end{equation}
where $M_P=1.22\times 10^{19}~{\rm GeV}$. We also have
\begin{equation}
\gamma _{_{S_{j}}}=2\gamma _{_{t_{j}}}^{(1)}+4\gamma
_{_{t_{j}}}^{(2)}.
\end{equation}%
The decay reaction density $\gamma _{_{D_{j}}}$ has the following
expression:
\begin{equation}
\gamma _{_{D_{j}}}=n_{N_{j}}^{eq}{\frac{K_{1}(z)}{K_{2}(z)}}\Gamma
_{N_{j}},
\end{equation}%
where $K_{n}(z)$ are the modified Bessel functions. $\Gamma
_{N_{j}}$ of the r.h.n $N_{j}$ is the tree level total decay rate
defined as
\begin{equation}
\Gamma _{N_{j}}=\frac{(\lambda^{\dagger} \lambda)_{jj}}{8\pi} M_{j},
\end{equation}%
where
\begin{equation}
n^{eq}_{N_i}(T)=\frac{g_i T
m_i}{2\pi^2}K_2\left(\frac{m_i}{T}\right).
\end{equation}
We used the definition $\lambda=M_D/\upsilon$. We define the reaction density $\gamma ^{(i)}$ of any process $%
a+b\rightarrow c+d$ by
\begin{equation}
\gamma ^{(i)}=\frac{M_{1}^{4}}{64\pi ^{4}}\frac{1}{z}\int_{\frac{%
(M_{a}+M_{b})^{2}}{M_{1}^{2}}}^{\infty }{dx~\hat{\sigma}^{(i)}(x)\sqrt{x}%
~K_{1}\left( \sqrt{x}z\right) },
\end{equation}%
where $\hat{\sigma}^{(j)}(x)$ are the reduced cross sections for the
different processes which contribute to the Boltzmann equations. For the $%
\Delta L=1$ processes involving the quarks, we have
\begin{eqnarray}
\hat{\sigma}_{t_{j}}^{(1)} &=&3\alpha _{u}\sum_{\alpha =1}^{3}\left(
\lambda
_{\alpha j}^{\ast }\lambda _{\alpha j}\right) \left( \frac{x-a_{j}}{x}%
\right) ^{2},  \label{sigma1} \\
\hat{\sigma}_{t_{j}}^{(2)} &=&3\alpha _{u}\sum_{\alpha =1}^{3}\left(
\lambda
_{\alpha j}^{\ast }\lambda _{\alpha j}\right) \left( \frac{x-a_{j}}{x}%
\right) \left[ \frac{x-2a_{j}+2a_{h}}{x-a_{j}+a_{h}}+\frac{a_{j}-2a_{h}}{%
x-a_{j}}\ln \left( \frac{x-a_{j}+a_{h}}{a_{h}}\right) \right],
\label{sigma2}
\end{eqnarray}%
where
\begin{equation}
\alpha _{u}=\frac{Tr(\lambda _{u}^{\dag }\lambda _{u})}{4\pi }\simeq \frac{%
m_{t}^{2}}{4\pi v^{2}},\hspace{0.5cm}a_{j}={\left( {\frac{M_{j}}{M_{1}}}%
\right) ^{2}},\hspace{0.5cm}a_{h}={\left( {\frac{\mu
}{M_{1}}}\right) ^{2}},
\end{equation}%
$\mu $ is the infrared cutoff which we set to $800~{\rm GeV}$
\cite{Luty,PlumacherSO10}.
For the $\Delta L=2$ r.h.n exchange processes, we have%
\begin{eqnarray}
\hat{\sigma}_{N}^{(1)} &=&\sum_{\alpha =1}^{3}\sum_{j=1}^{3}\left(
\lambda _{\alpha j}^{\ast }\lambda _{\alpha j}\right) \left( \lambda
_{\alpha
j}^{\ast }\lambda _{\alpha j}\right) A_{jj}^{(1)}+\sum_{\alpha =1}^{3}\sum_{%
{n<j},{j=1}}^{3}Re\left( \lambda _{\alpha n}^{\ast }\lambda _{\alpha
j}\right) \left( \lambda _{\alpha n}^{\ast }\lambda _{\alpha
j}\right)
B_{nj}^{(1)}  \label{sigmaN1} \\
\hat{\sigma}_{N}^{(2)} &=&\sum_{\alpha =1}^{3}\sum_{j=1}^{3}\left(
\lambda _{\alpha j}^{\ast }\lambda _{\alpha j}\right) \left( \lambda
_{\alpha
j}^{\ast }\lambda _{\alpha j}\right) A_{jj}^{(2)}+\sum_{\alpha =1}^{3}\sum_{%
{n<j,}{j=1}}^{3}Re\left( \lambda _{\alpha n}^{\ast }\lambda _{\alpha
j}\right) \left( \lambda _{\alpha n}^{\ast }\lambda _{\alpha
j}\right) B_{nj}^{(2)}  \label{sigmaN2}
\end{eqnarray}%
where%
\begin{eqnarray}
A_{jj}^{(1)} &=&\frac{1}{2\pi }\left[ 1+\frac{a_{j}}{D_{j}}+\frac{a_{j}x}{%
2D_{j}^{2}}-\frac{a_{j}}{x}\left( 1+\frac{x+a_{j}}{D_{j}}\right) \ln
\left(
\frac{x+a_{j}}{a_{j}}\right) \right] \mathit{,} \\
A_{jj}^{(2)} &=&\frac{1}{2\pi }\left[ \frac{x}{x+a_{j}}+\frac{a_{j}}{x+2a_{j}%
}\ln \left( \frac{x+a_{j}}{a_{j}}\right) \right] , \\
B_{nj}^{(1)} &=&\frac{\sqrt{a_{n}a_{j}}}{2\pi }\left[ \frac{1}{D_{j}}+\frac{1%
}{D_{n}}+\frac{x}{D_{j}D_{n}}+\left( 1+\frac{a_{j}}{x}\right) \left( \frac{2%
}{a_{n}-a_{j}}-\frac{1}{D_{n}}\right) \ln \left( \frac{x+a_{j}}{a_{j}}%
\right) \right. \\
&+&\left. \left( 1+\frac{a_{n}}{x}\right) \left( \frac{2}{a_{j}-a_{n}}-\frac{%
1}{D_{j}}\right) \ln \left( \frac{x+a_{n}}{a_{n}}\right) \right] ,\nonumber \\
B_{nj}^{(2)} &=&\frac{\sqrt{a_{n}a_{j}}}{2\pi }\left\{ \frac{1}{x+a_{n}+a_{j}%
}\ln \left[ \frac{(x+a_{j})(x+a_{n})}{a_{j}a_{n}}\right] +\frac{2}{%
a_{n}-a_{j}}\ln \left( \frac{a_{n}(x+a_{j})}{a_{j}(x+a_{n})}\right)
\right\} ,
\end{eqnarray}%
and
\begin{equation}
D_{j}={\frac{\left( x-a_{j}\right) ^{2}+a_{j}c_{j}}{x-a_{j}}},\hspace{1cm}%
c_{j}=a_{j}\sum_{\alpha =1}^{3}{\frac{\left( \lambda _{\alpha
j}^{\ast
}\lambda _{\alpha j}\lambda _{\alpha j}^{\ast }\lambda _{\alpha j}\right) }{%
64\pi ^{2}}}.
\end{equation}%
Finally, $\gamma _{_{W}}$ that accounts for the washout processes in
the Boltzmann equations is
\begin{equation}
\gamma _{_{W}}=\sum_{j=1}^{3}\left( \frac{1}{2}\gamma _{_{D_{j}}}+\frac{%
Y_{N_{j}}}{Y_{N_{j}}^{eq}}\gamma _{_{t_{j}}}^{(1)}+2\gamma _{_{t_{j}}}^{(2)}-%
{\frac{\gamma _{_{D_{j}}}}{8}}\right) +2\gamma _{N}^{(1)}+2\gamma
_{N}^{(2)}\ .  \label{gammaW}
\end{equation}
Here, we emphasize the so-called RIS (real intermediate states) in
the $\Delta L=2$ interactions which have to be carefully subtracted
to avoid double counting in the Boltzmann equations. This
corresponds to the term $-{\frac{1}{8}}\gamma _{_{D_{j}}}$ in Eq.
(\ref{gammaW}). For more details see Refs.
\cite{efficiency,underwood} and the first paper of Ref.
\cite{recentefficiency}.

\section{Results and discussion}

We are now ready to present our numerical results. First we make
several important remarks. Even though our model is supersymmetric,
we have considered in our BE analysis only the SM particle
interactions. This is a good approximation (see footnote 7). The
authors in Ref. \cite{PlumacherSO10}
 have demonstrated that SUSY interactions do not
significantly change the final baryon asymmetry. Furthermore, we
have not included in our analysis the effects of renormalization
group on the running masses and couplings. The first paper of Ref.
\cite{recentefficiency} has studied these effects. This paper has
also included finite temperature effects and $\Delta L=1$ scattering
processes involving SM gauge bosons, which we have ignored in our
analysis. This should be a good approximation since it is believed
that these effects are significant in the weak washout regime and
our model parameters seem to favor the strong washout regime with
$\widetilde{m}_1=\frac{(M^{\dagger}_D M_D)_{11}}{M_1}\simeq 0.1~{\rm
eV}$. Scattering processes involving gange bosons have also been
studied in Ref. \cite{underwood} in the context of resonant
leptogenesis where they have been shown to be significant.

Our next step is to put this model to the test and check its
predictions. In order to compute the value of the baryon asymmetry
we proceed to numerically solve the Boltzmann equations. We scan the
parameter space corresponding to the parameters $a_{12}$, $b$, the
oscillation angle $\theta_{13}$, the CP phase $\delta$ and the
Majorana phase $\alpha$.
In order to automatically satisfy the oscillation data, we input the
following light neutrino parameters:
\begin{equation}\label{data}
\Delta m^2_{\odot}= 2.5 \times 10^{-5}{\rm eV}^2,\,\,\Delta m^2_{a}=
5.54\times 10^{-3} {\rm eV}^2,\,\, \sin\theta_{12} = 0.52.
\end{equation}
Using hierarchical spectrum, we see that the masses $ m_1$, $m_2$
and $m_3$ are fixed. On the other hand we consider maximal mixing in
the 2-3 sector of the leptonic mixing matrix, i.e $\theta_{23}=
\frac{\pi}{4}+t_{23}\epsilon$ with $t_{23}$ being zero (
$t_{23}\sim{\cal O}(1)$ has minimal impact on $\eta_B$). The CP
phase $\delta$ and the Majorana phase $\alpha$ are allowed to vary
in the intervals $[0, 2\pi]$ and $[0, \pi]$ respectively. We remind
the reader that the second Majorana phase $\beta$ is related to
$\alpha$ through $\beta \simeq \alpha + \frac{\pi}{2}+b\epsilon$.
$\theta_{13}$ will be allowed to vary in the interval [0; 0.2] as it
is bounded from above by  reactor neutrino experiments.

In Figure (2), for a given set of input parameters, we illustrate
the different thermally averaged reaction rates
$\Gamma_{_X}={\gamma_{_X}\over n^{eq}_{N_1}}$ contributing to BE as
a function of  $z=\frac{M_1}{T}$.

All rates at $z=1$ fulfill the out of equilibrium condition (i.e.
$\Gamma_X\lesssim H(z=1)$), and so the expected washout effect due
to the $\Delta L=2$ processes will be small. The parameters chosen
for this illustration are: $\delta=\pi/2$, $\alpha=\pi/2$,
$a_{12}=0.01$, $b=0.9$,
$cm_{\tau}=m_t\left(\frac{m_{\tau}}{m_b}\right)=135~{\rm GeV}$ and
$\theta_{13}=0.02$. Eq. (\ref{data}) fixes the light neutrino masses
to be: $ m_{1}=0.00271292~{\rm eV}$, $m_2=0.00688186~{\rm eV}$ and
$m_3=0.0380442~{\rm eV}.$
 For this choice we obtain
$\mid\epsilon_1\mid\simeq 2 \times 10^{-7}$. The calculated r.h.n
masses in this case are
\begin{equation}\label{heavymasses}
M_1=9\times 10^9~{\rm GeV},~~~ M_2=8.7\times 10^{11}~{\rm GeV},~~~
M_3=2.6\times 10^{14}~{\rm GeV}.
\end{equation}
The mass of the lightest r.h.n is consistent with lower bound
derived in Ref. \cite{recentefficiency}, $M_1 \ge 2.4 \times 10^9$
GeV, for hierarchical neutrino masses assuming that one starts with
zero $N_1$ initial abundance (which is what we assumed in our
calculation). This mass is also in accordance with the upper bound
found in Ref. \cite{Davidson} following a model independent study of
the CP asymmetry, and the bound derived in Ref. \cite{efficiency}
based on the estimation of $\nu_R$ production and the study of the
asymmetry washout.

Figure (3) represents the solution of the BE, $N_1$ abundance and
the baryon asymmetry both as functions of $z$ for the same set of
parameters mentioned above. The final baryon asymmetry, in terms of
the baryon to photon ratio, is (see dark, solid curve in Fig. (3)
for $z\gg 1$)
\begin{equation}
\eta_{B} \simeq 6.03\times 10^{-10}.
\end{equation}
This number is inside the observational range of Eq. (\ref{wmap}).
Our codes were tested to reproduce the results in the first paper of
Ref. \cite{LeptoConstraints} before being applied to this model.

At this point we should mention some potential difficulties with
gravitino decays in the model.  If supersymmetry breaking is
mediated by conventional supergravity, it is natural to expect the
mass of the gravitino to be of order 1 TeV.  Such a gravitino, with
its Planck scale suppressed interactions, would decay into MSSM
particles with a lifetime of order 1 second.  The decay products
would upset the success of standard big bang nucleosynthesis.  In an
inflationary scenario, demanding consistency with light element
abundance puts a lower limit on the reheating temperature $T_R$. For
gravitino mass in the range $300$ GeV to 3 TeV, a limit $T_R > 6
\times 10^6$ GeV has been derived \cite{moroi}.  Now, the decays of
$N_1$ that generates lepton asymmetry should occur after reheating,
since any asymmetry produced prior to that will be diluted by
inflation.  Thus $M_1 < T_R$ is required, which for gravitino mass
in the range $300$ GeV to 3 TeV is in conflict with the predictions
of Eq. (\ref{heavymasses}).

There are several ways around this problem.  (i) In gauge mediated
SUSY breaking scenario the gravitino is the lightest SUSY particle
with mass in the range $10^{-4} ~{\rm eV}- 100 ~{\rm GeV}$.  For
$m_{\tilde{g}} < 100 $ MeV, there are no cosmological or
astrophysical problems.  In such a scenario the axion can serve as
the dark matter. (ii) In anomaly mediated SUSY breaking scenario,
the gravitino mass is enhanced by a loop factor compared to the
squark masses and is naturally of order 100 TeV.  Such a gravitino
would decay with a shorter lifetime without affecting big bang
nucleosynthesis. The gaugino is a natural dark matter candidate in
this case.  (iii) The gravitino itself can be the LSP and dark
matter with a mass of order 100 GeV, in which case it does not decay
\cite{feng}. Other solutions include changing the dynamics of the
leptogenesis process by invoking (iii) non--thermal leptogenesis
\cite{nonthermal}, (iv) resonant leptogenesis \cite{underwood,
johnellis}, or (v) soft leptogenesis \cite{softlepto}. The results
presented here are compatible with any one of the first three
alternatives.
\begin{figure}
\begin{center}
\epsfig{file=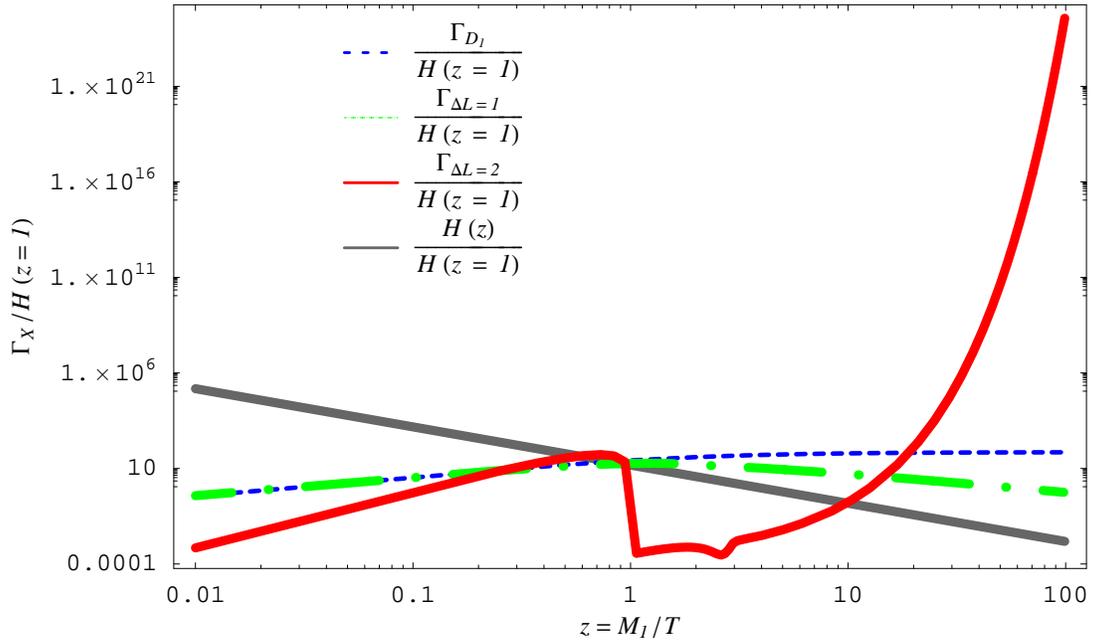,width=6.in} \caption{Various thermally
averaged reaction rates $\Gamma_X$ contributing to BE normalized to
the expansion rate of the Universe $H(z=1)$. The straight greyed
line represents $H(z)/H(z=1)$, the dashed is for
$\Gamma_{D_1}/H(z=1)$, the dotted-dashed line represents
$\Gamma_{\Delta L=1}/H(z=1)$ processes and the red curve represents
$\Gamma_{\Delta L=2}/H(z=1)$.}
\end{center}
\end{figure}
\begin{figure}
\begin{center}
\epsfig{file=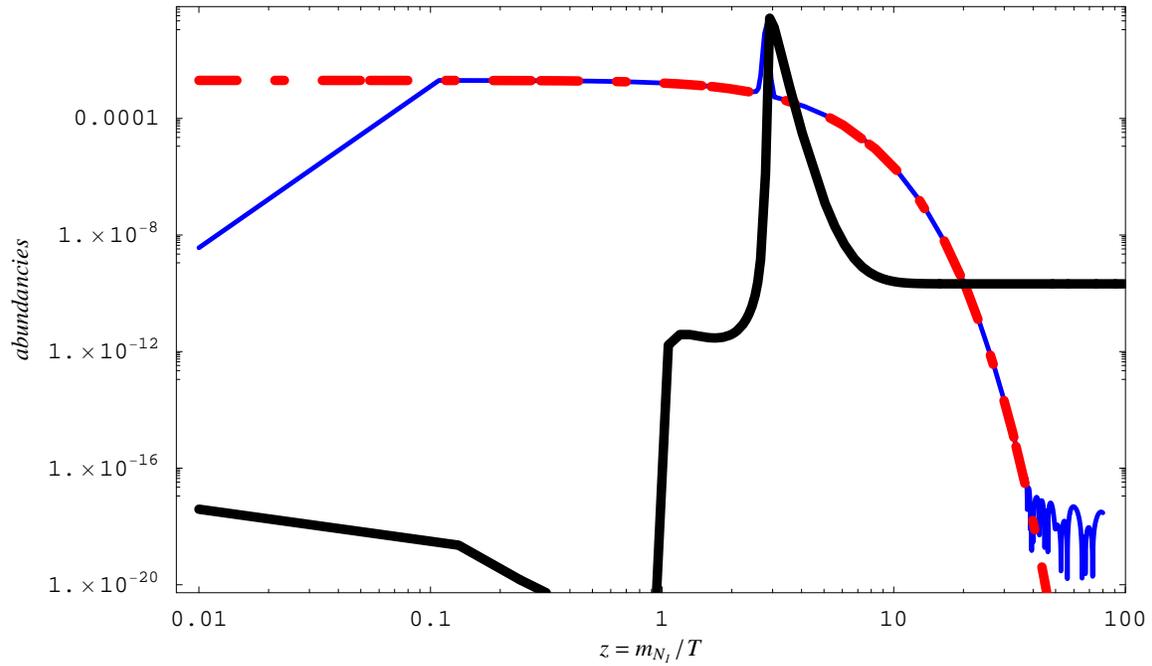,width=6.in} \caption{Evolution of $Y_{N_1}$
(solid blue), $Y_{N1}^{eq}$ (dot-dash) and the baryon asymmetry
$\eta_{B}$ (dark solid line) in terms of z in the model. The
estimated value for the baryon asymmetry is $\eta_{B} \simeq
6.03\times 10^{-10}$, with $Y_{N_1}^{ini}=0$ and assuming no
pre-existing $B-L$ asymmetry.}
\end{center}
\end{figure}
\newpage
\section{Conclusion}
An attractive feature of the seesaw mechanism is that it can explain
the origin of small neutrino masses and at the same time account for
the observed baryon asymmetry in the universe by the out of
equilibrium decay of the super-heavy right handed neutrinos. It is
then very tempting to seek a link between the baryon asymmetry
parameter $\eta_B$ induced at high temperature and neutrino mass and
mixing parameters observable in low energy experiments. No
quantitative connection can be found between them in the SM. There
have been several attempts in the literature \cite{LeptoConstraints,
lepto2, falcone} to establish a relationship between the two. In
this paper we have addressed this question in the context of a class
of minimal left--right symmetric models.

In the models under consideration the minimality of the Higgs sector
implies that $M_l$ and $M_D$ (charged lepton and Dirac neutrino mass
matrices) are proportional. As a result, the entire seesaw sector
(including the heavy right--handed neutrinos and the light
neutrinos) has only 9 parameters. This is the same number as low
energy neutrino observables (3 masses, 3 mixing angles and 3
phases). As a result we are able to link the baryon asymmetry of the
universe to low energy neutrino observables. This feature is unlike
the SM seesaw which has too many arbitrary parameters. Our numerical
solution to the coupled Boltzmann equations shows that this
constrained system with $M_l \propto M_D$ leads to an acceptable
baryon asymmetry. The requirement of an acceptable baryon asymmetry
restricts some of the light neutrino observables. We find that
$\tan^2\theta_{12} \simeq m_1/m_2$, $0.01\lesssim
\theta_{13}\lesssim 0.07$ and  $\beta \simeq \alpha + \pi/2$  are
needed for successful baryogenesis. Future neutrino oscillation
experiments can directly probe into the dynamics of the universe in
its early stages.
\section{Acknowledgements}
One of the authors (A. B) wishes to thank I. Gogoladze for
discussions and A. Strumia for useful correspondence. The work of K.
B and A. B is supported in part by US Department of Energy grant
\#DE-FG02-04ER46140 and \#DE-FG02-04ER41306.

\end{document}